\documentclass[prc,twocolumn,showpacs,preprintnumbers,amsmath,amssymb]{revtex4}
\usepackage{graphicx}

\usepackage{dcolumn}
\usepackage{bm}
\begin{document}
\bibliographystyle{/usr/share/texmf/tex/latex/revtex/prsty}
\draft

\title{Realistic two-nucleon  potentials for the relativistic two-nucleon
Schroedinger equation}

\author{
H. Kamada
\footnote{email:kamada@mns.kyutech.ac.jp}
}

\address{
Department  of Physics, Faculty of Engineering, Kyushu Institute of Technology,
 Kitakyushu 804-8550, Japan}

\author{
W. Gl\"ockle
\footnote{email:walter.gloeckle@tp2.ruhr-uni-bochum.de}
}

\address{
Institut f\"ur  Theoretische Physik II,
Ruhr-Universit\"at Bochum, 44780 Bochum, Germany}

\date{\today}
\begin{abstract}
 The potentials $V ( v)$ in the nonrelativistic ( relativistic) nucleon-nucleon (NN) Schr\"odinger
equation are related by a quadratic equation. That equation is numerically solved, thus
providing phase equivalent $v $ - potentials related for instance  to the  high precision NN
potentials, which are  adjusted to NN phase shift and mixing parameters in a nonrelativistic
Schr\"odinger equation. The relativistic NN potentials embedded in a three-nucleon (3N)
  system for total NN
momenta different from zero are also constructed in a numerically precise manner. They enter
into the relativistic interacting 3N mass operator, which is needed for relativistic 3N
calculations for bound and scattering states.
\end{abstract}

\maketitle

\title{Realistic two-nucleon  potentials for the relativistic two-nucleon
Schroedinger equation}

\section{Introduction}

  Traditionally the (semi) phenomenological high precision two-nucleon
  (NN)potentials AV18 \cite{AV18},
 CD Bonn \cite{CDBONN}
 and Nijm I,II \cite{NIJM}
 go together with
  the nonrelativistic operator for the kinetic energy $ \frac{\hat{k}^2}{m}$ in the
  NN c.m.system. Nevertheless, as is well known,  this nonrelativistic
  Schr\"odinger equation
\begin{eqnarray}
(\frac{\hat{k}^2}{m} + V ) \Psi = \frac{k_0^2}{m} \Psi\label{1}
\end{eqnarray}
can be related to an underlying relativistic NN Schr\"odinger
equation
\begin{eqnarray}
( 2 \sqrt{m^2 + \hat{k}^2} + v) \Phi = 2 \sqrt{m^2 + k_0^2} \Phi\label{2}
\end{eqnarray}
by a simple algebraic step\cite{CPS},\cite{friar}. Applying $( 2 \sqrt{m^2 +
\hat{k}^2} + v)  $ to (\ref{2}) from the left one obtains
\begin{eqnarray}
( 4 ( m^2 + \hat{k}^2 ) + 2 \omega(\hat{k}) v + 2 v \omega(\hat{k}) + v^2 ) \Phi = 4 ( m^2
+ k_0^2) \Phi\label{3}
\end{eqnarray}
which can be identically rewritten into (\ref{1}) if one defines
\begin{eqnarray}
V = \frac{1}{4m} ( 2 \omega(\hat{k}) v + 2 v \omega(\hat{k}) + v^2)\label{4}
\end{eqnarray}
with $ \omega(\hat{k}) = \sqrt{m^2 + \hat{k}^2}$. ( We use $\hat{k}$ in order to distinguish the
momentum operator from the number $k$).  Therefore adjusting $ V $ in
(\ref{1}) to the NN phase shift and mixing parameters from a phase
shift analysis and relating the c.m. momentum $ k_0$ to the
Lorentz invariant lab energy $ T_{lab}$ via
\begin{eqnarray}
k_0 = \sqrt{\frac{m T_{lab}}{2}}\label{5}
\end{eqnarray}
( a relation identically valid for relativistic and nonrelativistic
kinematics) one has in fact solved a relativistic equation. We
also see that $ \Psi $ equals $ \Phi$ .The question remains, what
is $ v $ given $ V$? The formal solution of that quadratic
equation (\ref{4}) is
\begin{eqnarray}
v = \sqrt{4mV + 4 \omega(\hat{k})^2} - 2 \omega(\hat{k})\label{6}
\end{eqnarray}
Why is $ v $ of interest? If one turns to the 3N system and would
like to investigate relativistic effects \cite{gloeckle86},\cite{witala05} the knowledge of $ v $
is very useful. It defines together with the relativistic kinetic
energy the interacting NN mass operator, which is a key ingredient
for building the interactive 3N mass operator \cite{coester65}. Therefore we
focus in this paper on the determination of $ v$ related to the
high precision NN potentials via (\ref{4}) or (\ref{6}).

In \cite{carlson93},\cite{forest95}
 a potential $ v$ has been determined directly fitting (\ref{2}) to
NN phase shifts. Thereby the Urbana $v_{14}$ potential has been
readjusted achieving a fair fit( though not of the quality of the high precision potentials).
In \cite{KG98} a momentum scale transformation
\begin{eqnarray}
2m + \frac{k^2}{m} = 2 \sqrt{m^2 + q^2}\label{7}
\end{eqnarray}
has been introduced which provides an analytical relation between
$ V $ and $ v$ and guarantees that the S-matrix related to (\ref{1}) at
c.m.momentum k equals the S-matrix related to (\ref{2}) at c.m. momentum
q. In other words the relativistic and nonrelativistic S-matrices
agree at the same energy. This,however, is misleading since the
equality of the two S-matrices should hold at the same c.m.momenta
\cite{allen00}.
 A better, though still approximate
 approach to relate $ V $
and $ v$ has been given in \cite{gloeckle86}. 

On the other hand there is the possibility to add
an interaction to the square of the free NN mass operator 
$ h^2 \equiv ( 2 \omega(\hat{k}) )^2 + 4m V~' $.
Then $ H \equiv \frac{h^2}{4m} -m $ has exactly the same form as (\ref{1}) with $ V~' = V$
 provided we identify
$h^2 $ with the mass operator of the interacting NN system \cite{ CPS}, \cite{gloeckle86}.
 This is of course also obvious from
the relations (\ref{1}) - ( \ref{4}).
The construction of the relativistic 3N Hamiltonian requires, however, the 3N mass operator $h$
rather than its square. Therefore 
our aim here is to solve (\ref{4}) and (\ref{6}) exactly for $ v$. This is
outlaid in section II. The validity of the resulting $ v$ is
 verified by demonstrating that it provides exactly the same
phase shift parameters using (\ref{2}) as the underlying $ V $ using (\ref{1}).

Next we  regard the 3N mass operator where NN c.m. forces enter in
the form \cite{coester65}
\begin{eqnarray}
v_p \equiv \sqrt{( 2 \omega(\hat{k}) + v)^2 + p^2} - 
\sqrt{( 2 \omega(\hat{k}))^2 +
p^2 }\label{8}
\end{eqnarray}
with $ p$ the total NN momentum. The $ p$-dependence arises since
in a 3-body system the NN subsystems are not at rest.
In section  III we propose a simple manner to determine $ v_p$ in
a numerical precise way.
 This opens now the door to use $ v's$ which are equivalent to
 the underlying high precision potentials in a relativistic
 context in 3N bound and scattering problems.
We summarize in section IV. A technical derivation is given in the Appendix.

\section{The potential {\bf ${\small v} $} }

The determination  of $ v$ using Eq. (\ref{6}) can be achieved by a
spectral decomposition. One can proceed in close analogy to the representation derived in 
\cite{kamada02} for $ v_p$ given in Eq. (\ref{8}). We regard a specific partial wave state (
or coupled ones) with given orbital angular momentum(a) , total
spin and total angular momentum. For the sake of a simpler
notation we will not show these quantum numbers explicitely. Using
the completeness relation of bound and scattering states for the
potential $ V$ one obtains
\begin{eqnarray}
&&<  k \vert v \vert  k~ '>= <  k \vert \Psi_b> M_b< \Psi_b \vert  k ~' >\cr
&&- 2 \omega(k) \frac{\delta( k - k')}{k k~'} \cr
&&+  \int_0^\infty d  k '' (k~'')^2 <  k \vert \Psi_ k'' > 2 \sqrt{
{k''^2 } +m^2 }< \Psi_ k''  \vert k ~'>
\cr &&
\label{9}
\end{eqnarray}
where $\Psi_b (k)$ is the  nonrelativistic deuteron wave function of
(\ref{1}) and  $M_b$ the mass of the deuteron. Here we introduced a different definition of the
binding energy, namely an implicit one:
\begin{eqnarray}
M_b \equiv \sqrt{4 m^2 + 4 \epsilon_b~' m} = 2m + \epsilon_b~' - \frac{{\epsilon_b~'}^2}{4m} + \cdots
\end{eqnarray}
In lowest order it agrees with the usual one $ M_b \equiv 2m + \epsilon_b $. This new definition
of the binding energy has in addition the feature that it can naturally be written as
 $ \epsilon_b~' = -\frac{\kappa^2}{m}$ in agreement with the form of the energy eigenvalue
 of (\ref{1}) at the bound
state pole $k= i \kappa $ of the S-matrix.

The expression (\ref{9}) can be identically rewritten into the form
\begin{eqnarray}
&&\langle  k \vert v  \vert  k' \rangle
 =  \Psi_b( k)   M_b  \Psi_b (k')
 \cr &+& {   m \over { k ^2 - {k '}^2 } } \{  2 \omega(k) \Re
[ T ( k' , k ;
 { {k}^2 \over m})]\cr & -& 
   2 \omega(k~')' \Re [ T ( k ,  k' ; { {k'}^2 \over m} )] \}
 \cr &+&  {m^2 \over {  k ^2 - {k '}^2 } }\times
 \cr
 & \{ &  {\cal P}  \int_0^\infty d k'' ( k~'')^2  { 2  \omega(k~'')  \over {{ { k ''} ^2 - {k }^2 }} }\cr
  & & T ( k ,  k'' ; { {k''}^2 \over m})
  T^{*} (k' ,  k'' ;  { {k''}^2 \over m} )
   \cr &-& {\cal P}
 \int_0^\infty d k'' (k~'')^2 {  2  \omega(k~'')  \over { { k ''} ^2 - {k '}^2 } }\cr
 & &   T ( k ,  k'' ;{ {k''}^2 \over m} )
   T^{*} ( k' ,  k'' ; { {k''}^2 \over m} )
    \}.  \label{11}
    \end{eqnarray}
  where $T$ is the standard NN T-matrix related to $V$ via the
    nonrelativistic Lippmann Schwinger equation.
The derivation of that form is defered to the Appendix.
A numerical implementation has not yet been performed, but we expect no problem.

A second more simple way is to directly solve the quadratic
operator equation (\ref{4}). In momentum space it reads
\begin{eqnarray}
& & 2 m < k| V | k~'>  = (\omega(k) + \omega(k~') )<k | v | k~'>\cr
& + &  \frac{1}{2} \int_0^\infty dk~'' {k~''}^2
< k|  v | k~''> < k~''| v | k~'>\label{12}
\end{eqnarray}
or
\begin{eqnarray}
& & <k | v | k~'>\cr
& + &  \frac{1}{2(\omega(k) + \omega(k~'))} \int_0^\infty dk~'' {k~''}^2
< k|  v | k~''> < k~''| v | k~'>\cr
& = &  \frac{2m < k| V | k~'>}{\omega(k) + \omega(k~')}\label{13}
\end{eqnarray}

 We verified numerically that for all the realistic high
 precision potentials AV18, CD Bonn, and Nijm I,II the following
 very simple iterative scheme works
 \begin{eqnarray}
<k | v | k~'>^{(0)} = \frac{2m < k| V | k~'>}{\omega(k) + \omega(k~')},
\label{13}
 \end{eqnarray}
\begin{eqnarray}
&& <k | v | k~'>^{(n+1)} = 
\frac{1}{2(\omega(k) + \omega(k~'))}
\times
\{ 4m < k |V|{k~'}>  
\cr
&&-
\int_0^\infty dk~'' {k~''}^2
< k|  v | k~''>^{(n)}< k~''| v | k~'>^{(n)}
\}
\label{14}
\end{eqnarray}
with $ n=0,1,\cdots$.
 For certain partial waves the iteration to converge requires an
 additional step, namely
 \begin{eqnarray}
 v^{(n+1)} \xleftarrow{\rm redefine} 
( a  v^{(n+1)}  + b v^{(n)}  ) /(a+b)\label{15}
\label{ab}
 \end{eqnarray}
where the constants $a$ and $b$ are typically 1.

\begin{figure}[tt]
\includegraphics[width=.300\textheight]{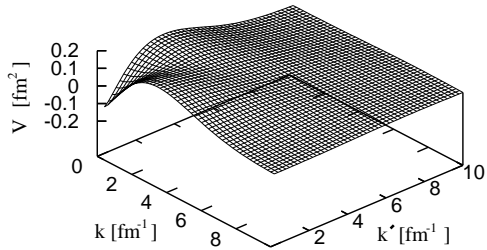}
\caption{The nonrelativistic potential $V(k,k')$ AV18 in the state $^1S_0$. \label{fig.1} } 
\end{figure}

\begin{figure}[tt]
\includegraphics[width=.300\textheight]{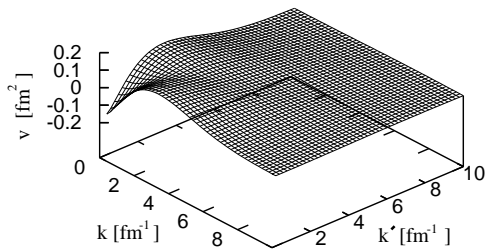}
\caption{ The relativistic potential $v(k,k')$ related to AV18 in the state $^1S_0$.\label{fig.2} } 
\end{figure}

\begin{figure}[tt]
\includegraphics[width=0.300\textheight]{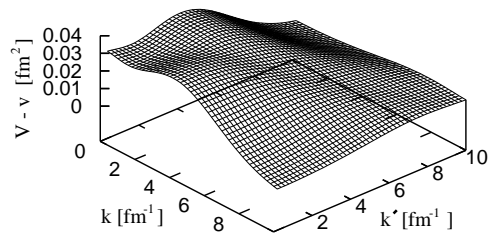}
\caption{The difference between the relativistic and nonrelativistic potentials: 
$V(k,k')-v(k,k')$ in the state  $^1S_0$. \label{fig.3} }
\end{figure}

 We display in Table \ref{tab:1} an example documenting the convergence. In
 Figs \ref{fig.1}-\ref{fig.3} we show  for an example the original
 nonrelativistic potential $ V (k,k~ ') $, the resulting
 relativistic potential $ v(k,k~ ')$ and the difference $ V(k,k~ ') -
 v(k,k~') $. We see in that example that $V(k,k~') > v(k,k~')$.


To further characterize the difference between $ V(k,k~')$ and $v(k,k~')$ one can regard the
asymptotic behavior 
\begin{eqnarray}
\lim_{k \to \infty, k~' {\rm fixed}} v(k,k~') = \frac{\rm const}{k^n}
\end{eqnarray}
against
\begin{eqnarray}
\lim_{k \to \infty, k~' {\rm fixed}} V(k,k~') = \frac{\rm const}{k^N}
\end{eqnarray}
As examples , for AV18 we find $n= 3$, $N= 2$, whereas CD Bonn delivers 
$n=4.5$, $N=3.5$.
 The value of $N$ for
CD Bonn can easily be understood \cite{M89}: $ N=2+2+1/2-1 $ with the two $2$'s resulting from the
meson propagator and the choice of the strong form factor, the  $ 1/2$ arising from transforming
the Blankenbeclar-Sugar equation to the nonrelativistic Lippmann Schwinger equation and the
$(-1)$ from the two Dirac spinors. In both cases, AV18 and CD Bonn, 
$n$ is larger than $N$ by
 one unit, which is suggested by Eq.(\ref{12}).

 \begin{table}
 \caption{Convergence of $v^{(n)}$ to the iteration in Eq. (\ref{14}). \\ 
We choose the coupled partial waves ($^3S_1$-$^3D_1$) of 
the Argonne V18 potential\cite{AV18}. The momenta $k$ and $k'$
are 1.0 fm$^{-1}$ and the potential unit is [fm$^2$]. 
 \label{tab:1}}
 \begin{ruledtabular}
 \begin{tabular}{ccccc}
        $n$       & $v^{(n)}$($^3S_1$-$^3S_1$) & 
                  $v^{(n)}$($^3S_1$-$^3D_1$) & 
                  $v^{(n)}$($^3D_1$-$^3D_1$) \\ \hline
0 &  0.084232 & 0.044709 & 0.016853 \\
1 & 0.067716  & 0.044628 & 0.016785 \\
2 & 0.059933  & 0.044597 & 0.016744 \\
3 & 0.056135  & 0.044587 & 0.016719 \\
4 & 0.054234  & 0.044585 & 0.016705 \\
5 & 0.053259  & 0.044587 & 0.016696 \\
6 &  0.052749  & 0.044589 & 0.016691 \\
10 & 0.052194  & 0.044595 & 0.016684 \\
20 & 0.052126  & 0.044597 & 0.016684 \\
30 & 0.052126  & 0.044597 & 0.016684 
 \end{tabular}
 \end{ruledtabular}
 \end{table}

Having $ v $ at our disposal one can solve the relativistic
Lippmann Schwinger equation

\begin{eqnarray}
& & t(k,k~ ') =  v(k,k~ ')\nonumber\\
&  + &  \int_0^\infty dk~ '' {k~ ''}^2 v(k,k~ '') \cr
& & \frac{1}{ 2 \omega( k~ ' ) - 2 \omega(k~ '') + i \epsilon } t( k~ '',k~ ')\label{16}
\end{eqnarray}
for the half shell t-matrix. It is related to the S-matrix via
\begin{eqnarray}
s(k) = e^{2 i \delta_r(k)} = 1 - i \pi k \omega(k) t(k,k).
\end{eqnarray}
The corresponding nonrelativistic relation is
\begin{eqnarray}
S(k) = e^{2i \delta_{nr}(k)} = 1 - i \pi k m T(k,k).
\end{eqnarray}
 where $ T(k,k~ ')$ obeys the standard nonrelativistic Lippmann
 Schwinger equation. We illustrate  in Table \ref{tab:2} for the  ($ ^3 S_1$ - $^3 D_1$ ) 
 partial wave states
 the perfect agreement of the corresponding
  phase shift and mixing parameters.

 \begin{table}
 \caption{ Comparison of the phase shift and the mixing parameters  \\
for  the coupled partial waves ($^3S_1$-$^3D_1$) of 
the Argonne V18  potential\cite{AV18}. 
The second column points to the  relativistic (Rel.) or 
nonrelativistic (Nonrel.) calculations.
The unit of the phases are  in 
degrees.
 \label{tab:2}}
 \begin{ruledtabular}
 \begin{tabular}{ccccc}
$T_{lab.}$ [MeV] & Nonrel./Rel. & $\delta$($^3S_1$) & 
                              $\delta$($^3D_1$) & 
                  $\epsilon$  \\ \hline
1.0   & Nonrel. &  147.62   & -0.0050743 & 0.10303 \\
1.0   & Rel.    &  147.62   & -0.0050744 & 0.10304 \\
10.0  & Nonrel. &  102.71   & -0.66593 & 1.1267  \\
10.0  & Rel.    &  102.71   & -0.66593 & 1.1267  \\
50.0  & Nonrel. &  62.929   & -6.3189  & 2.0853  \\
50.0  & Rel.    &  62.929   & -6.3189  & 2.0854  \\
100.0 & Nonrel. &  43.531   & -12.093  & 2.4899  \\
100.0 & Rel.    &  43.531   & -12.093  & 2.4899  \\
350.0 & Nonrel. &  2.6451   & -26.719  & 4.9223  \\
350.0 & Rel.    &  2.6452   & -26.719  & 4.9222  \\
 \end{tabular}
 \end{ruledtabular}
 \end{table}

\section{The construction of the NN force embedded in the  3N system}

\begin{figure}[tt]
\includegraphics[width=.300\textheight]{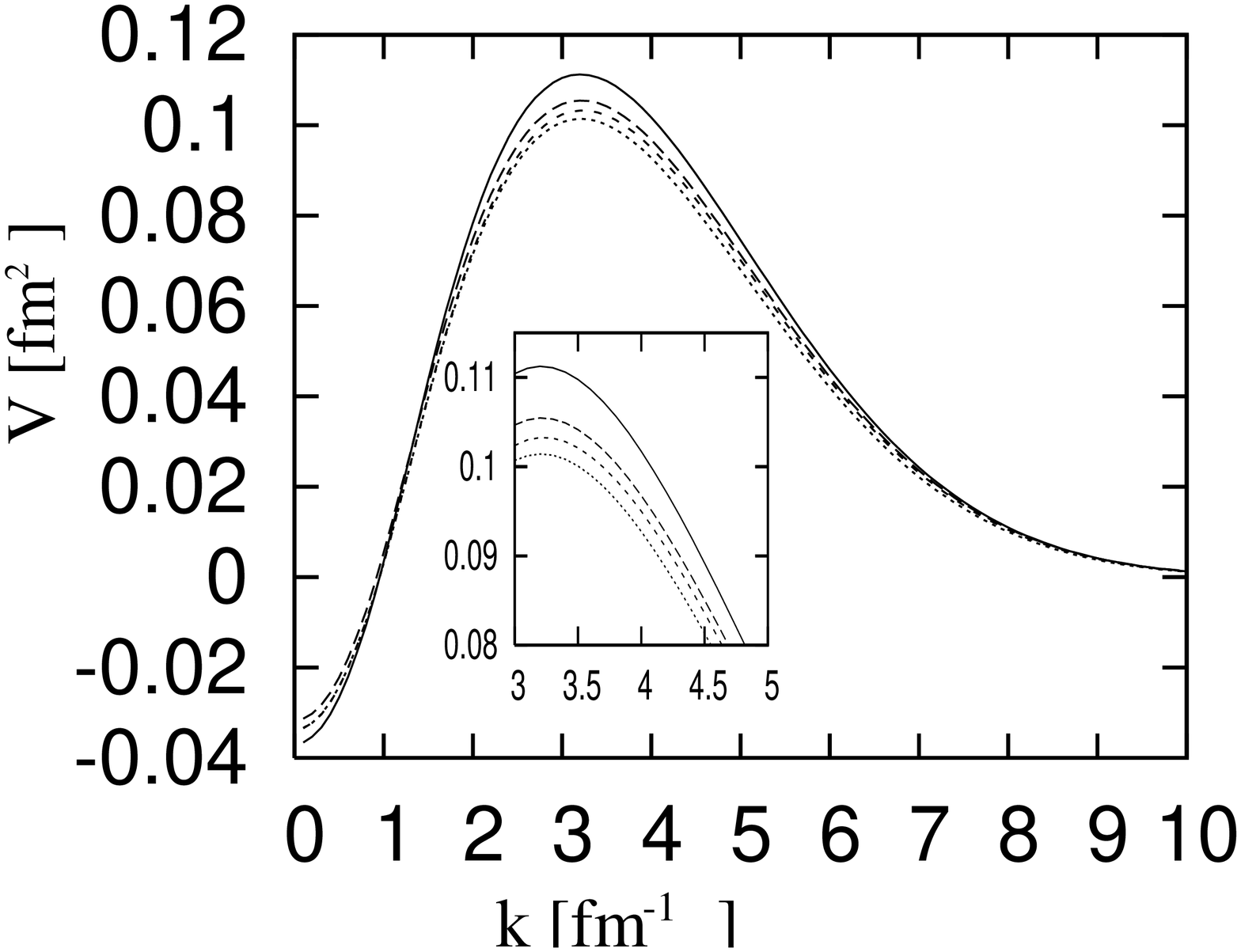}
\caption{Comparison $v( k,k~')$ against $v_p( k,k~') $ for AV18 in the state  $^1S_0$
for fixed  $k'=1.0$ fm$^{-1}$.
The solid curve shows the relativistic potential $v$  ($p=0$).
The other curves show $v_p$ for  $p=4.0$fm$^{-1}$.
The long dashed curve is the exact one, the short dashed and dotted curves show
 the approximations for $ v_p$ given in Eqs.(\ref{approx1}) and (\ref{approx2}), respectively.
\label{fig.4} } 
\end{figure}

If one defines $ 2 \omega_p(\hat{k}) \equiv \sqrt{ 4 (\omega(\hat{k}))^2 + p^2}$ then
using (\ref{4}) Eq. (\ref{8}) can be written 
\begin{eqnarray}
& & ( v_p + 2 \omega_p(\hat{k}))^2 = ( 2 \omega(\hat{k}) + v )^2 + p^2\cr
& = &  4 (\omega_p(\hat{k}))^2 + 4mV
\end{eqnarray}
or 
\begin{eqnarray}
(v_p)^2 + 2 v_p \omega_p(\hat{k}) + 2 \omega_p(\hat{k}) v_p = 4m V
\end{eqnarray}
Between free states it yields
\begin{eqnarray}
& & v_p(k,k~') ( \omega_p(k) + \omega_p(k~'))\cr
&  = &  2m V(k,k~')\cr
&  - &  \frac{1}{2}\int_0^\infty d k~'' ( k~'')^2
v_p( k,k~'') v_p(k~'',k~')
\end{eqnarray}
This has the same structure as (\ref{13}) replacing $\omega(k)$ by $ \omega_p(k)$. The iteration
procedure described in section II works equally well. 
We compare in Fig \ref{fig.4} $ v(k,k~')$ and
 $ v_p(k,k~')$.  We see a  weakening of $ v_p $ against $ v$. This is a fact known from previous
calculations \cite{kamada02} and from the approximate ( but very useful) expression \cite{ witala05}
\begin{eqnarray}
v_p(k,k~') = v(k,k~') ( 1- \frac{p^2}{ 8 \omega(k) \omega(k~')})
\label{approx1}
\end{eqnarray}
 This is demonstrated in Fig \ref{fig.4}, where also an even simpler approximation
\begin{eqnarray}
v_p(k,k~') = v(k,k~') ( 1- \frac{p^2}{ 8 m^2})
\label{approx2}
\end{eqnarray}
is shown. That latter approximate version, however, is somewhat worse and is not recommended.

\section{Summary}
Relativistic calculations in the instant form of dynamics proposed in \cite{Dirac} requires an
interacting 3N mass operator. As has been shown in \cite{coester65} and used in
\cite{gloeckle86},\cite{witala05},\cite{kamada02},  \cite{lin},
\cite{HW06}, and \cite{SK06} the NN potential
 in a moving frame $ v_p$ enters in the form
given in Eq.(\ref{8}), where the NN force in the NN c.m.system $ v$ enters into the
relativistic NN Schr\"odinger equation (\ref{2}). We showed that the quadratic operator
relation (\ref{4}) for $v$ can be solved directly in an iterative manner and this very
precisely. This has been documented by evaluating NN phase shift and mixing parameters using
the standard nonrelatvistic Schr\"odinger equation (\ref{1}) and the relativistic NN
Schr\"odinger equation (\ref{2}). This opens the way to get $ v$'s related to any NN potential $V$
adjusted in a nonrelativistic frame work like the high precision NN potentials. By the same
iterative procedure also $ v_p$ can be gained. It turned out that $ v $ is smaller in
magnitude than $ V$ and $ v_p$ is smaller than $v$. Applications to the 3N bound and scattering
states are planned.

The numerical calculations has been performed on the IBM Regatta p690+ of the NIC in J\"ulich, Germany.

\section{Appendix}

Derivation of Eq.(\ref{11})

\noindent
We start from (\ref{9}) and  use the well known decomposition
\begin{eqnarray}
 <  k \vert \Psi_k' > = \frac{\delta ( k - k')}{k k~'}+ 
{ T( k, k'; {k'^2 \over m }) \over {k'^2 \over m }+ i \epsilon -  {k^2
 \over m } }
\label{a1}
\end{eqnarray}
to arrive at
\begin{eqnarray}
&&<  k \vert v \vert  k~ '>=
 \Psi_b ( k)  M_b \Psi_b ( k ~')  \cr
&+& { T^*( k', k; {k^2 \over m }) \over {k^2 \over m }- i \epsilon -
  {k'^2 \over m } } 2 \omega( k) \cr
&+& { T( k,k'; {k'^2 \over m }) \over {k'^2 \over m }+
 i \epsilon -  {k^2 \over m } } 2 \omega( k') \cr
&+&\int_0^\infty d  k '' ( k~'')^2  { T( k, k''; {k''^2 \over m }) \over {k''^2 \over m }
+ i \epsilon -  {k^2  \over m } } 2 \omega( k'')\cr
& &   { T^*( k',k''; {k''^2 \over m }) \over {k''^2 \over m }-
 i \epsilon -  {k'^2  \over m } }
\label{a2}
\end{eqnarray}
The integral requires some care and we keep the limiting processes for
the two scattering states separately by putting
\begin{eqnarray}
&&
{1 \over k''^2 - k ^2 +i \epsilon }
{1 \over k''^2 - {k' } ^2 -i \epsilon } \cr &\to&
{1 \over k''^2 - k ^2 +i \epsilon_1 }
{1 \over k''^2 - {k' } ^2 -i \epsilon_2 }
\cr
&=& \left( {1 \over k''^2 - k ^2 +i \epsilon_1 } -
{1 \over k''^2 - {k' } ^2 -i \epsilon_2 } \right) \times
\cr &&
{1 \over k^2 - {k' } ^2 -i( \epsilon_1 + \epsilon_2)  } .
\label{a4}
\end{eqnarray}
This allows us to perform one limit firstly with the result
\begin{eqnarray}
&&
{1 \over k''^2 - k ^2 +i \epsilon }
{1 \over k''^2 - {k' } ^2 -i \epsilon } \cr &\to&
\left( { {\cal P  } \over   k''^2 - k ^2 } - i \pi \delta ( k''^2 - k
^2 ) \right) {1 \over k^2 - {k' } ^2 -i \epsilon_2  } \cr
&-&
\left( { {\cal P  } \over   k''^2 - {k' } ^2 } + i \pi \delta ( k''^2 -{k' }
^2 ) \right) {1 \over k^2 - {k' } ^2 -i \epsilon_1  }
\label{a5}
\end{eqnarray}
Thus we get for some well behaved function $f ( k '') $
\begin{eqnarray}
&&
\int_0^\infty d k~''(k~'')^2 { f ( k '') \over ( k''^2 - {k } ^2 +i
\epsilon_1 )
(k''^2 - {k' } ^2 -i \epsilon_2)  } \cr
&&
= \lim _{\epsilon \to +0  } {1 \over k^2 - {k' } ^2 -i \epsilon } \times
\cr &&
\left( {\cal P } \int_0^\infty d  k '' (k~'')^2  {f( k '' ) \over k''^2 - k^2  }
-  {\cal P } \int_0^\infty d  k '' ( k~'')^2  {f( k '' ) \over k''^2 -{ k'}^2
}
\right) \cr
&&- i\pi \lim_{\epsilon \to +0  } {1 \over k^2 - {k' } ^2 -i \epsilon } \times
\cr
 && ( \int_0^\infty d  k '' (k~'')^2  f( k '' ) \delta ( k''^2 - k^2 )\cr
& + &  \int_0^\infty d  k ''(k~'')^2  f( k '' ) \delta (k''^2 -{ k'}^2 ) )
\label{a6}
\end{eqnarray}
The principal value prescription is denoted as ``${\cal P}\int$''.
In our case
\begin{eqnarray}
f( k '') =
2m^2 \omega (k'') T( k,  k'' ; { k''^2 \over m }) T^*( k',
 k'' ; { k''^2 \over m })
\label{a7}
\end{eqnarray}
and therefore
\begin{eqnarray}
&&\int_0^\infty d  k '' (k~'')^2 f( k'') \delta ( k''^2 - k^2 ) \cr
&=&2m^2 \int_0^\infty d  k '' ( k~'')^2 \omega(k'')  T( k,  k'' ; { k''^2
\over m })\cr
& &  T^*( k', k'' ; { k''^2 \over m }) \delta ( k''^2 - k^2 )\cr
& = & m^2 k \omega(k) T(k,k; { k^2 \over m }) T^*( k', k ; { k^2 \over m })
\label{a8}
\end{eqnarray}
This is part of the unitary relation
\begin{eqnarray}
& &  T( k, k' ; { q^2 \over m }) - T^*( k',
 k ; { q^2 \over m })\cr
 &=&  2i \Im T(k,k~'; { q^2 \over m })\cr
& = & 2i \Im T(k~',k; { q^2 \over m })\cr
& - & 2 i \pi m \int_0^\infty d  k''( k~'')^2  T( k,  k'' ; { q^2 \over m }) T^*( k',
 k'' ; { q^2 \over m }) \delta( q^2 - k''^2)\cr
& = & - i \pi m q T(k,q;{ q^2 \over m }) T^*( k~',q;{ q^2 \over m })
\cr &&
\label{a9}
\end{eqnarray}
( Note we used the symmetry of the off- the- energy- shell T-matrix). Consequently
\begin{eqnarray}
&&  \int_0^\infty d k''  k~''^2 f(k~'')   \delta (k''^2 - k^2)
\cr &=&
- {2 \omega(k) m \over \pi  }
\Im [ T( k ,  k' ; { k ^2 \over m } ) ]
\label{a10}
\end{eqnarray}
and
\begin{eqnarray}
&& -i\pi \lim_{\epsilon \to +0  } { 1 \over k^2 - {k' }^2  -i \epsilon} \times
\cr &&
( \int_0^\infty d  k '' (k~'')^2 f(k'') \delta (k''^2 - k^2)\cr
& +& \int_0^\infty d  k''(k~'')^2 f(k'') \delta ( k''^2 - {k' }^2 ))\cr
&&=\lim _{\epsilon \to +0  } { 2 m i \over k^2 -{k'}^2-  i \epsilon } \times
\cr &&
\left(
\omega (k) \Im [T (  k, k' ; { k^2 \over m  } )] +
\omega (k') \Im [T ( k~',  k ; { {k'}^2 \over m })]  \right)
\label{a11}
\end{eqnarray}
Combined with Eq. (\ref{a2}) certain terms cancel and one arrives at
Eq.(\ref{11}).

\end{document}